\documentclass[aps,prd,twocolumn,preprintnumbers,superscriptaddress,nofootinbib]{revtex4-1}
\usepackage{graphicx}
\usepackage{epstopdf}
\usepackage{amsmath}
\usepackage{amsfonts}
\usepackage{amssymb}
\usepackage{appendix}
\usepackage{enumerate}
\usepackage{natbib}
\usepackage{comment}
\usepackage{bbold}
\usepackage[shortlabels]{enumitem}
\usepackage{color}
\usepackage{slashed}
\usepackage{subfigure}
\usepackage{setspace}
\usepackage{lipsum}
\usepackage{multirow}
\usepackage[colorlinks = true,
            linkcolor = blue,
            urlcolor  = blue,
            citecolor = blue,
            anchorcolor = blue]{hyperref}
\usepackage[capitalize]{cleveref}
\usepackage{braket}
\usepackage[compat=1.1.0]{tikz-feynman}
\usepackage{multirow}
\usepackage{physics}
\usepackage{feynmp-auto}
\usepackage[normalem]{ulem}
\usepackage{url}
\usepackage{units}
\usepackage[normalem]{ulem}

\def\bar{\overline}

\begin{document}

\singlespacing

\preprint{FERMILAB-PUB-23-063-T}
\preprint{\hspace{0.15cm}NUHEP-TH/23-01, \hspace{0.15cm} CERN-TH-2023-024}
\preprint{\hspace{0.15cm}USTC-ICTS/PCFT-23-06}

\title{The Neutrino Magnetic Moment Portal and Supernovae:\\ New Constraints and Multimessenger Opportunities}

\author{Vedran Brdar} 
\email{vedran.brdar@cern.ch}
\affiliation{Theoretical Physics Department, CERN,
Esplande des Particules, 1211 Geneva 23, Switzerland}
\author{Andr\'{e} de Gouv\^{e}a} 
\email{degouvea@northwestern.edu}
\affiliation{Northwestern University, Department of Physics \& Astronomy, 2145 Sheridan Road, Evanston, IL 60208, USA}
\author{Ying-Ying Li}
\email{yingyingli@ustc.edu.cn}
\affiliation{Peng Huanwu Center for Fundamental Theory, Hefei, Anhui 230026, China}
\affiliation{Interdisciplinary Center for Theoretical Study,
University of Science and Technology of China, Hefei, Anhui 230026, China}
\author{Pedro A.~N.~Machado}
\email{pmachado@fnal.gov}
\affiliation{Particle Theory Department, Fermilab, P.O. Box 500, Batavia, IL 60510, USA}

\begin{abstract}
We scrutinize the hypothesis that gauge singlet fermions - sterile neutrinos - interact with Standard Model particles through the transition magnetic moment portal. These interactions lead to the production of sterile neutrinos in supernovae followed by  their decay into photons and active neutrinos which can be detected at $\gamma$-ray telescopes and neutrino detectors, respectively. We find that the non-observation of active neutrinos and photons from sterile-neutrino decay associated to SN1987A yields the strongest constraints to date on magnetic-moment-coupled sterile neutrinos if their masses are inside a $0.1-100$~MeV window. Assuming a near-future galactic supernova explosion, we estimate the sensitivity of several present and near-future experiments, including Fermi-LAT, e-ASTROGAM, DUNE, and Hyper-Kamiokande, to magnetic-moment-coupled sterile neutrinos. We also study the diffuse photon and neutrino fluxes produced in the decay of magnetic-moment coupled sterile neutrinos produced in all past supernova explosions and find that the absence of these decay daughters yields the strongest constraints to date for sterile neutrino masses inside a $1-100$~keV window. 
\end{abstract}

\maketitle

\textit{\bf{Introduction ---}}
While the Standard Model (SM) shows remarkable consistency with numerous experiments, it fails to account for nonzero neutrino masses, dark matter, and the baryon asymmetry of the Universe. The addition of gauge-singlet fermions -- sterile neutrinos -- to the SM allows one to address these shortcomings. Sterile neutrinos allow for tiny active neutrino masses via the type-I seesaw scenario \cite{Minkowski,Goran,GellMann:1980vs,Yanagida:1979as}, a model that can also dynamically generate the baryon asymmetry via leptogenesis \cite{Fukugita:1986hr,Akhmedov:1998qx}. Further, light sterile neutrinos have been widely discussed as viable dark matter candidates \cite{Drewes:2016upu}. In the absence of more degrees of freedom, at the renormalizable level, a sterile neutrino $N$ can only interact with SM particles through Yukawa interactions. These are constrained at different levels by a variety of experimental probes across a plurality of mass scales \cite{Bolton:2019pcu,Abdullahi:2022jlv}. 

At the non-renormalizable level, sterile neutrinos can also interact with neutrinos and photons through a magnetic-moment-type interaction. In the SM, neutrino magnetic moments are expected to be very small \cite{Fujikawa:1980yx, Lee:1977tib, Petcov:1976ff, Pal:1981rm, Shrock:1982sc, Dvornikov:2003js, Giunti:2014ixa, Tanabashi:2018oca} but they can be enhanced in beyond the Standard Model (BSM) scenarios, particularly those associated with the origin of neutrino masses \cite{Voloshin:1987qy, Barbieri:1988fh, Babu:1989wn, Babu:1989px, Lindner:2017uvt, Xu:2019dxe, Babu:2020ivd,Alok:2022pdn}. Likewise, in the presence of sterile neutrinos, large active-to-sterile neutrino transition magnetic moments can be generated as a consequence of more BSM physics \cite{Brdar:2020quo,Magill:2018jla,Schwetz:2020xra,Ismail:2021dyp,Smirnov:2022suv,Brdar:2022rhc}. Some are related to the existence of new TeV-scale new physics motivated by other anomalies in particle physics (see, e.g., \cite{Brdar:2020quo}). 

Active-to-sterile neutrino transition magnetic moments are described by a Lagrangian that includes, after electroweak symmetry breaking,
\begin{align}
    \mathcal{L} \supset \sum_\alpha d_\alpha \bar{N}\sigma_{\mu\nu} \nu^{\alpha} F^{\mu\nu}-\frac{M_N}{2} \bar{N}^c N + \text{h.c.}\,.
    \label{eq:Lag}
\end{align}
Here, $\nu^{\alpha}$ ($\alpha = e, \mu, \tau$) are the active neutrino fields while $F^{\mu\nu}$ is the field strength tensor of the electromagnetic field. $d_\alpha$ are the interaction strengths, with units of inverse energy; for simplicity, we assume flavor universal interactions and define $d_\alpha\equiv d, \forall\alpha$. $M_N$ is the mass of the sterile neutrino. Throughout, we assume that $N$ and the active neutrinos are Majorana fermions.

Given the large flux of neutrinos associated to core-collapse supernova explosions (SN) \cite{Mirizzi:2015eza}, sterile neutrinos can be efficiently produced via magnetic-moment interactions; the process $\nu e^-\to N e^-$ is the dominant production channel for $M_N \lesssim 100$ MeV \cite{Magill:2018jla}. The observed properties of the SN1987A neutrinos, including their total energy \cite{Loredo:2001rx,Pagliaroli:2008ur,Huedepohl2010} and the estimated cooling time \cite{Kamiokande-II:1987idp,Bionta:1987qt,Baksan} -- around 10~s -- imply that hypothetical sterile neutrinos produced in SN1987A did not carry away  $\mathcal{O}(1)$ of the available energy. 
This energy loss argument has been widely employed in the literature in order to constrain various BSM scenarios, including keV-scale sterile neutrinos interacting through active--sterile mixing \cite{Raffelt:2011nc,Arguelles:2016uwb,Suliga:2020vpz}, axion-like particles \cite{Lucente:2021hbp,Caputo:2022rca} (see also complementary study \cite{Caputo:2022mah}), and dark photons \cite{Kazanas:2014mca,DeRocco:2019njg}.  For the model in \cref{eq:Lag}, a detailed analysis based on the energy loss argument was carried out in \cite{Magill:2018jla} and allows one to exclude $10^{-13}\, \text{MeV}^{-1} \lesssim d \lesssim 10^{-10}\, \text{MeV}^{-1}$ for $M_N \lesssim 100$ MeV \cite{Magill:2018jla} (see also \cite{PhysRevD.100.083002}). The lower bound on $d$ was set assuming the new interaction is such that sterile neutrinos carry at most 10\% of the energy released in the explosion, as is typically done in the literature. 

\cref{eq:Lag} also mediates the decay of the sterile neutrinos, $N\to \nu\gamma$, which potentially lead to observable signatures at neutrino detectors and $\gamma$-ray telescopes. We explore such multimessenger signatures and derive new limits and sensitivity projections. These turn out to be stronger than the standard energy-loss bounds discussed above when $M_N$ is in the $0.01-100$ MeV range.

\begin{figure}[t]
	\centering
	\includegraphics[scale=0.55]{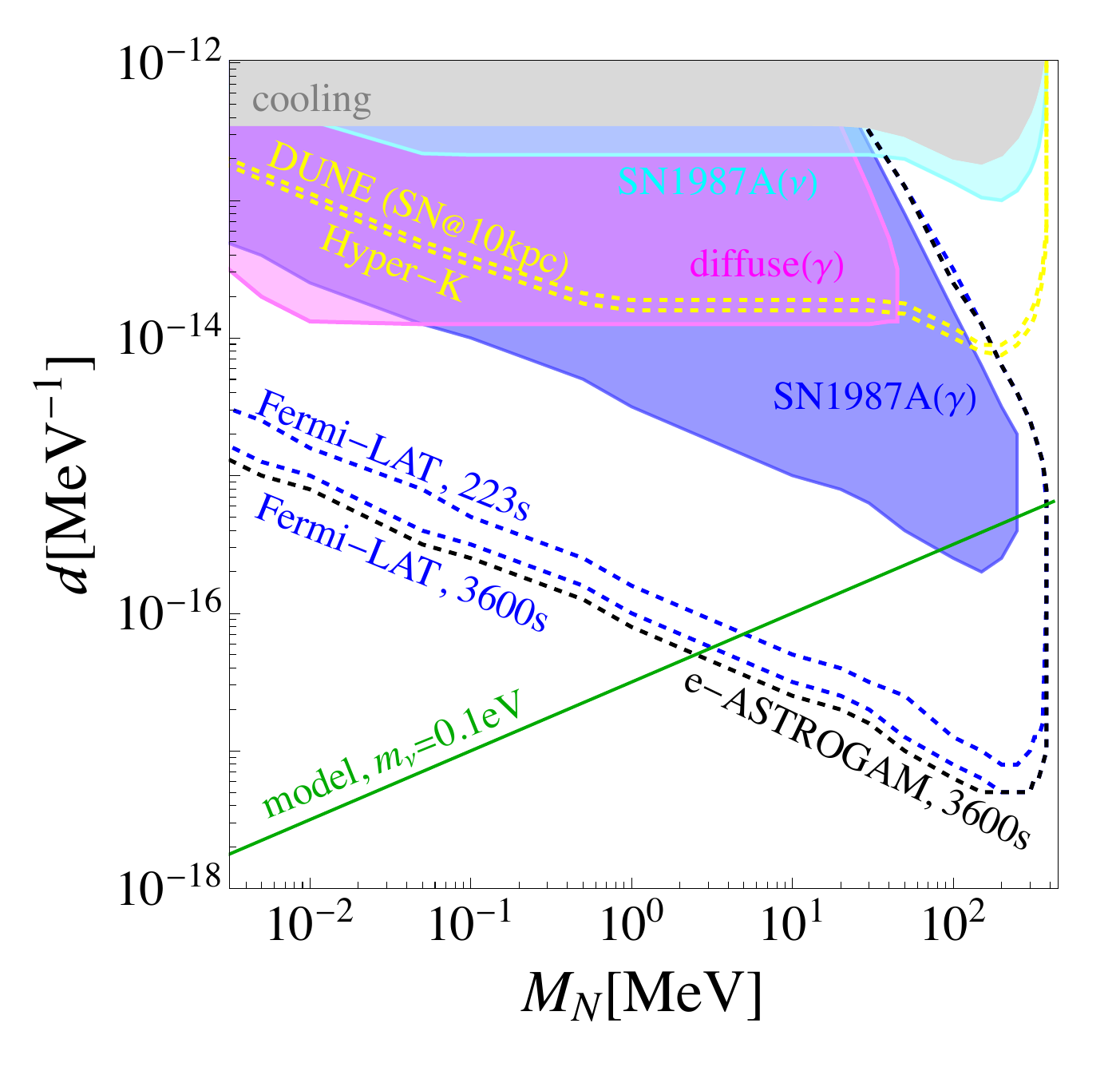} 
	\caption{2$\sigma$ constraints and future sensitivity for the transition magnetic moment $d$ as a function of sterile neutrino mass $M_N$. }
	\label{fig:exclusion}
\end{figure}

\textit{\bf{Sterile neutrino production ---}} 
The differential number of sterile neutrinos $\mathcal{N}_s$ produced via the magnetic moment portal per unit time $t$ at position $r$ is \cite{Arguelles:2016uwb}
\begin{align}
\frac{1}{4\pi r^2}\frac{\partial^2}{\partial r\partial t}\left(\frac{d\mathcal{N}_s}{dE_N}\right)= \sigma n_e \frac{d n_\nu}{dE}\,.
\label{eq:simplified}
\end{align}
Here, $\sigma n_e$ is related to the interaction rate for the process $\nu e^- \to N e^-$ which dominates, for $M_N \lesssim 100$ MeV, due to large neutrino and electron number densities inside the star. The cross section, $\sigma$, is proportional to $d^2$ and includes finite temperature effects as well as those of Pauli blocking on the final state electron; the latter means that a time and radius-dependent chemical potential has also been employed. Further, $n_e$ and $n_\nu$ are, respectively, the number densities of electrons and neutrinos. $(1/n_\nu)\, d n_\nu/dE=f(E)/\bar{E}$ where $\bar{E}$ is the mean neutrino energy and $f(E)$ is the neutrino distribution function \cite{Brdar:2018zds}. The approximate relation between the sterile-neutrino and active-neutrino energy is $E = \frac{1}{2}(E_N + p_N)$, where $p_N$ is the sterile-neutrino momentum. We solve \cref{eq:simplified} using data associated to the simulation performed by the Garching group of an $8.8 M_\odot$ progenitor star \cite{Huedepohl2010}. For more details, see Appendix \ref{app:integrate}. In addition to $\nu e^-$ scattering, we also include contributions from the inverse decay process, $\gamma \nu \to N$, which is known to be relevant when $M_N \gtrsim 100~\rm MeV$~\cite{Magill:2018jla}. Requiring the total energy carried away by $N$ to be less than $10\%$ of the total available neutrino energy leads to the  ``cooling'' bound depicted in \cref{fig:exclusion} (gray region). Our estimate is in agreement with the one obtained in \cite{Magill:2018jla}. 

While active neutrinos exit the SN with energies of a few tens of MeV, they are copiously produced at higher temperatures in the dense SN core. Sterile neutrinos, instead, in the parameter space of interest, promptly exit the SN after production. This implies that the typical sterile-neutrino energies are $\mathcal{O}(100~\rm MeV)$ (see \cref{fig:fluxes} in Appendix \ref{app:integrate}). Sterile neutrinos will decay en route to the Earth via $N\to \nu \gamma$, producing $\mathcal{O}(100~\rm MeV)$ active neutrinos and photons.

\begin{figure}[t]
	\centering
	\includegraphics[scale=0.08]{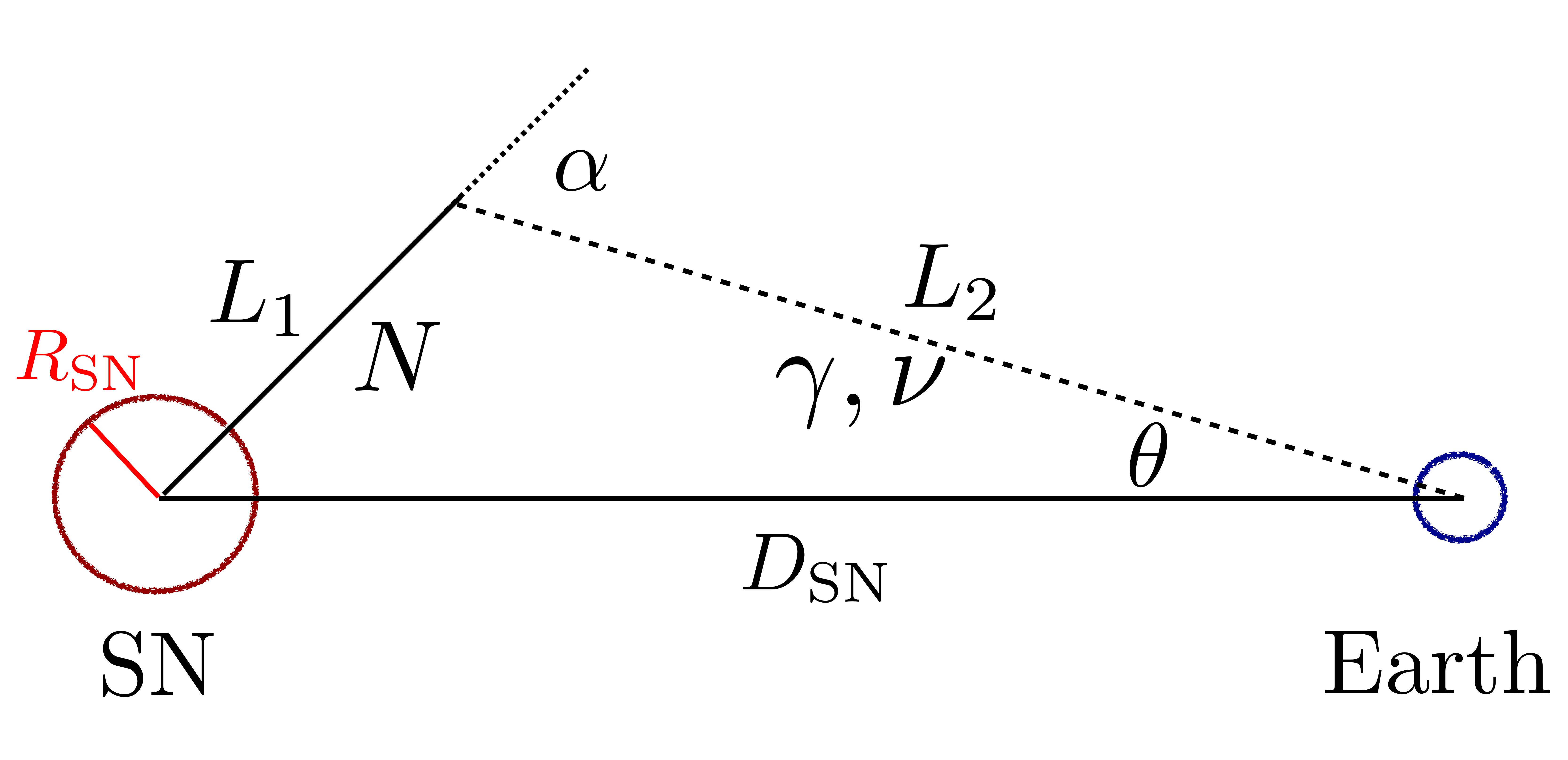} 
	\caption{Sterile neutrino decay geometry.}
	\label{fig:geometry}
\end{figure}

\textit{\bf{Sterile neutrino decay ---}}
The decay width for $N\to\nu\gamma$ is $\Gamma_N = 6d^2 M_N^3/4 \pi$ \cite{Plestid:2020vqf}. For $d$ values of interest, sterile neutrinos produced in the SN core do not interact before decaying. The daughter particles are emitted at an angle $\alpha$ relative to the direction of $N$, given by
\begin{align}
    \cos\alpha &= \frac{2 E_N E_{\gamma/\nu} - M^2_N}{2E_{\gamma/\nu}\sqrt{E^2_N-M^2_N}}\,,
\end{align}
where $E_\gamma$ and $E_\nu$ denote the energies of the daughter photon and neutrino, respectively.
For a small range of $\alpha$, the daughter particle can reach the Earth, as sketched in \cref{fig:geometry}.
The angle $\theta$ at which the daughter particle arrives at the Earth is given by $D_{\rm SN} \sin\theta = L_1 \sin\alpha$, where $D_{\rm SN}$ is the distance between the SN and the Earth, and $L_1$ is the distance propagated by $N$ before decaying, see \cref{fig:geometry}.

In the frame of an observer at the Earth, summing over all final-state photon and neutrino polarizations, the differential decay rate of a Majorana $N$ is given by the box distribution \cite{BahaBalantekin:2018ppj, BALANTEKIN2019488, Plestid:2020vqf} 
\begin{equation}
    \frac{d\Gamma}{d E_{\gamma/\nu}} = \Gamma_N \left(\frac{M_N}{E_N}\right) \frac{\Theta(E_{\gamma,\nu} - E^{-})\,\,\Theta(E^{+} - E_{\gamma,\nu})}{E^{+} - E^{-}}\,,
\label{eq:box}
\end{equation}
where $E^{\pm} = E_N\left(1\pm\beta\right)/2$ and $\beta = \sqrt{E^2_N-M^2_N}/E_N$, the speed of $N$.  
The time delay $\Delta t$, the arrival time of the daughter $\gamma$/$\nu$ relative to that of the neutrinos produced in the explosion via SM processes, is \cite{Jaeckel:2017tud}
\begin{align}
    \Delta t = L_1/\beta +L_2 -D_{\rm SN}\,,
    \label{eq:deltat}
\end{align}
with $L_2 = D_{\rm SN} \cos\theta-L_1 \cos\alpha$. For a given $\Delta t$, $L_1 \leq \beta \Delta t /(1-\beta)$. One finds that $\theta$ is bounded: $\sin\theta \leq L_1/d_{\rm SN} \leq \beta \Delta t /[(1-\beta) d_{\rm SN}$.

The flux of daughter $\gamma/\nu$ is obtained by integrating over all $N$ decays that occur at $R^{\gamma/\nu}_{\rm SN}\leq L_1 \leq L^{\rm max}_1$. 
$L^{\rm max}_1$ corresponds to the distance $L_1$ associated to the largest time delay considered or the largest angle $\theta_{\rm max}$ allowed by observations. $R^{\gamma/\nu}_{\rm SN}$ is the smallest decay distance for which the decay daughter can escape the explosion unperturbed. 
For daughter photons, $N$ should decay beyond the photosphere, otherwise the associated photon flux is severely attenuated. 
Conservatively, we take the photosphere radius to match the radius of the star; we adopt $R^{\gamma}_{\rm SN} = 3\times \unit[10^{10}]{ m}$, the estimated radius of the SN1987A progenitor \cite{prog}. For neutrinos, $R^{\nu}_{\rm SN}$ is chosen to be the radius of the neutrinosphere, assumed to be $R^{\nu}_{\rm SN}=30$~km. 
In what follows, we will consider different observation time windows and take $\theta_{\rm max}$ to be the angular resolution of the $\gamma$-ray  telescope.
With the sterile neutrino decay length $L_N= (E_N/M_N)\Gamma_N^{-1} \beta$, the differential flux of daughter particles per unit energy and area $A$ at the Earth is
\begin{align}
    \frac{d^2N_{\gamma/\nu}}{dE_{\gamma/\nu}dA} =\int\frac{e^{-R^{\gamma/\nu}_{\rm SN}/L_N}-e^{-L^{\rm max}_1/L_N}}{4\pi D^2_{\rm SN}(E^{+} - E^{-})}\frac{d\mathcal{N}_s}{dE_N}\, dE_N\,.
    \label{eq:flux_gamma_nu}
\end{align}

\begin{figure}[t]
	\centering
	\includegraphics[scale=0.55]{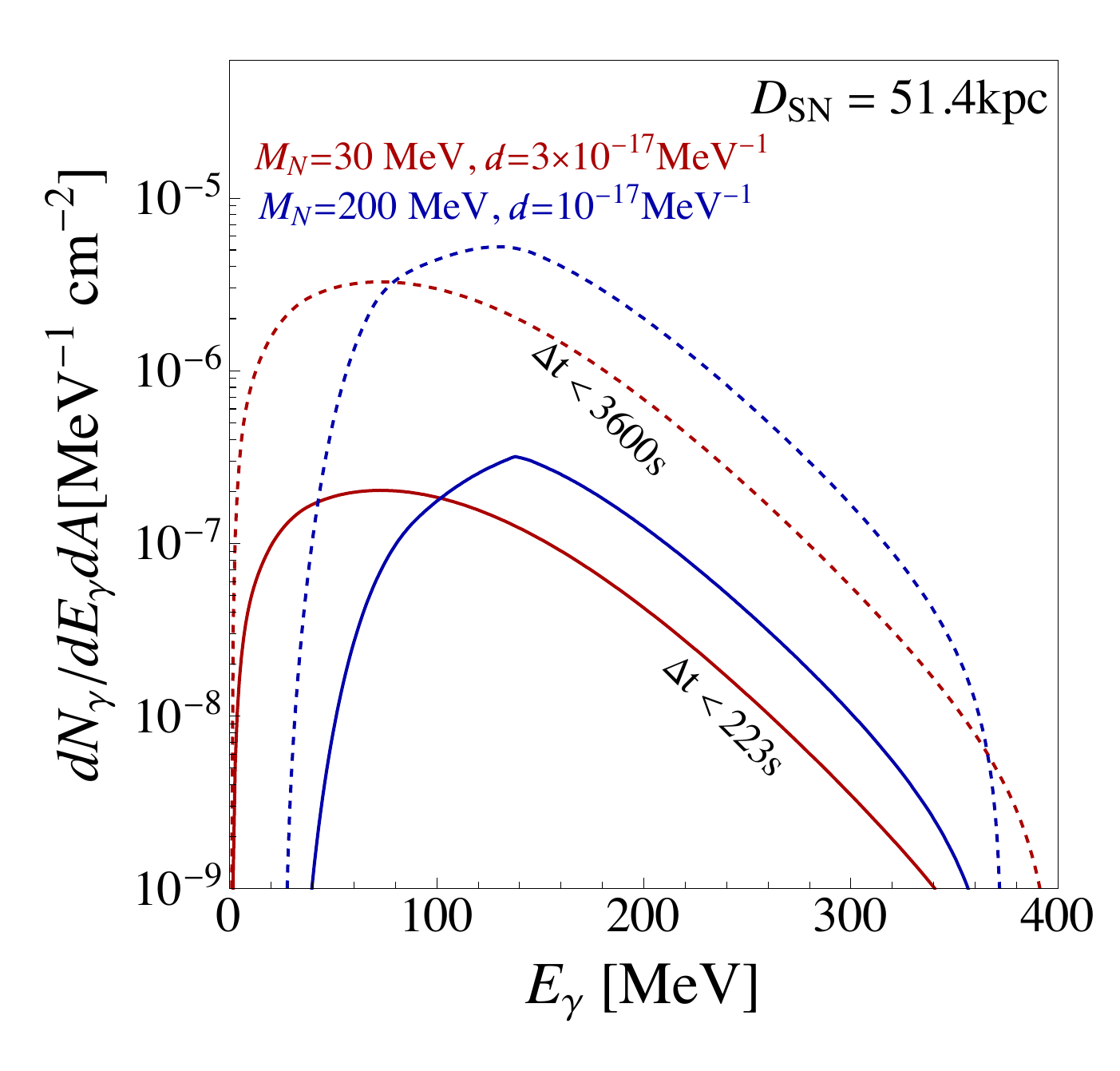} 
	\caption{Photons flux (at the Earth) from sterile neutrino
	decay, for $\Delta t <223$~s (solid) and $\Delta t <3600$~s (dashed).}
	\label{fig:photon_flux}
\end{figure}

\textit{\bf{Gamma-ray detection ---}}
We calculate the photon flux at the Earth using \cref{eq:flux_gamma_nu}. Our results are depicted in \cref{fig:photon_flux} for a time window $\Delta t < 223$~s (solid), and a nominal longer exposure $\Delta t < 3600$~s (dashed),  assuming $D_{\rm SN} =\unit[51.4]{kpc}$. 
We choose $\theta_{\rm max} = 5^{\circ}$ which, for $\Delta t < 3600$ s, always exceeds the angles at which the photons arrive. 
The ``plateau'' for $E_\gamma \lesssim M_N$ is a reflection of the aforementioned box energy distribution while suppression at larger $E_\gamma$ occurs because the production of sterile neutrinos with energies above the SN core temperature is inefficient. 
The total number of observed photons $N_\gamma^{\text{BSM}}$, including detection inefficiencies, is obtained by integrating \cref{eq:flux_gamma_nu} over the photon energy range and the detection area of the $\gamma$-ray telescope of interest.

Keeping in mind the small number of decay-daughter gamma rays and neutrinos from the SN explosion, we define the log likelihood
\begin{align}
-2{\rm ln}\mathcal{L} = 2 \left(N_\text{exp}-N_\text{obs}+N_\text{obs}\,\, \text{Log}\left[N_\text{obs}/N_\text{exp}\right] \right)\,,
\label{eq:poissonian}
\end{align}
and consider constraints at the $2\sigma$ level, associated with $-2{\rm ln}\mathcal{L}=3.841$. Here, $N_\text{exp}$ is the number of expected events 
for a particular point in the ($d\times M_N$) parameter space and $N_\text{obs}$ is the number of observed events.

At the time of SN1987A, the Gamma-Ray Spectrometer (GRS), mounted on the satellite-borne Solar Maximum Mission (SMM) \cite{SMM}, was operating and observed $N_\text{obs} = 1393$ events from the direction of SN1987A in the time window up to 223 s after the neutrino burst and in the energy band 25-100 \text{MeV}, with a full field view of the detector \cite{OBERAUER1993377,Caputo:2021rux}. This event number is consistent with expectations from the galactic diffuse photon flux and several other sources, including scintillation light induced by primary cosmic rays \cite{SMM}. We use $\theta_{\rm max} = 5^{\circ}$ to calculate the number of BSM events, $N^{\rm BSM}_\gamma$, in the time window $\Delta t < 223$~s. In the parameter space of interest, the angular distribution is very narrow ($\theta \ll 5^{\circ}$). Using \cref{eq:poissonian}, we require $N^{\rm BSM}_\gamma \le 76$, a 2$\sigma$ fluctuation of $N_\text{obs} = 1393$.  
The corresponding constraint is depicted in \cref{fig:exclusion} as a blue shaded region\footnote{We have neglected the possibility of fireball formation \cite{DeRocco:2019njg, Diamond:2023scc} in which case, a small part of the parameter space inside the blue shaded region would be constrained by Pioneer Venus Orbiter mission \cite{ Diamond:2023scc} instead of SMM.} labeled ``SN1987A$(\gamma)$''. For $M_N\sim 100$ MeV, we can exclude $d\approx 10^{-15}\, \text{MeV}^{-1}$, a constraint that significantly exceeds those of other probes, including the energy loss argument discussed earlier (gray region in \cref{fig:exclusion}).
For small $M_N$, the constraints are much weaker because the $\Delta t<223$~s condition becomes increasingly more difficult to satisfy.

In the event of a future core-collapse SN, current and near future experiments will be able to probe a much larger region of the parameter space.
Assuming the SN event happens in the galaxy at a distance $D_{\rm SN}=\unit[10]{kpc}$, which is not unlikely~\cite{Reed:2005en,Rozwadowska:2020nab},  
we consider the currently-operating Fermi-LAT, which has a total detection area of $\unit[9500]{cm^2}$ and angular resolution around $5^\circ$ in all directions \cite{Fermilat}, together with near-future experiments such as e-ASTROGAM \cite{e-ASTROGAM:2017pxr} (detection area $\unit[9025]{cm^2}$ and angular resolution of $1.25^\circ$), ComPair \cite{Moiseev:2015lva}, and PANGU \cite{Wu:2014tya}, which will provide better sensitivity for $E_\gamma\lesssim\unit[100]{MeV}$. e-ASTROGRAM, ComPair, and PANGU may be relevant for the small $M_N$ region, where a potentially significant portion of the flux has $E_\gamma<100$ MeV; see \cref{fig:photon_flux}, where fluxes for low and high $M_N$ benchmarks are compared.
To estimate the sensitivity of these experiments to transition magnetic moments, we consider the same time windows as before, 223~s and 3600~s. 
With these choices, for $E_N$ and $M_N$ values of interest, the arrival angle $\theta$ is, roughly, less than $5^\circ$.

When estimating the sensitivity of Fermi-LAT, we restrict $E_\gamma>\unit[100]{MeV}$ \cite{Fermi-LAT:2014ryh}.
For e-ASTROGAM, we instead consider photons with $E_\gamma>1$~MeV. 
The sensitivity projections are depicted in \cref{fig:exclusion} as dashed lines for both Fermi-LAT, with $\Delta t < 223\,\text{s}\,(3600\,\text{s})\,$ and $N^{\rm BSM}_\gamma < 2.5 \,(4.9)$ (blue); and e-ASTROGAM with $\Delta t < 3600\,\text{s}$ and $N^{\rm BSM}_\gamma < 4.4$ (black). These correspond to $2\sigma$ sensitivity, assuming all backgrounds can be eliminated. 
Future sensitivity is expected to improve on SN1987A constraints by roughly two orders of magnitude throughout the parameter space. 
PANGU will likely be able to measure the polarization of the gamma rays, allowing one to, for example, distinguish photons from different origins \cite{Jaeckel:2017tud, Balaji:2019fxd, Balaji:2020oig}. We leave this to a future study. 

To appreciate the fundamental physics impact of possible future Fermi-LAT and e-ASTROGAM measurements, we compute the transition magnetic moment predicted by a specific UV-complete model of neutrino masses involving leptoquarks~\cite{Brdar:2020quo}. There, active neutrino masses $m_\nu$ are obtained via the type-I seesaw mechanism. The expected values of the transition magnetic moment $d$ are related to the mechanism that generates the Dirac masses, connecting the active and sterile neutrinos. One obtains $d\simeq 10^{-13}$\, $\text{MeV}^{-2} \sqrt{m_{\nu} M_N}$ for leptoquark masses at the TeV scale. For $m_\nu= \unit[0.1]{eV}$, the expected value of $d$ as a function of $M_N$ is depicted in \cref{fig:exclusion} as a green line, well within the reach of the projected sensitivity for the next galactic SN event. 


\textit{\bf{Neutrino detection ---}}
Besides the photon signal, the production of $N$ can also lead to a new, higher energy SN neutrino flux.
For active neutrinos produced in radiative $N$ decay, we investigate two water-Cherenkov detectors, Kamiokande-II and IMB, that recorded, respectively, 11 and 8 neutrino events from SN1987A. In these experiments, antineutrinos were detected through inverse beta decay, the most relevant interaction channel for $\bar{\nu}_e$ detection in the $E_\nu \sim 10-50$~MeV window. 
Since we assume the magnetic moments are flavor universal, radiative $N$ decays lead to identical numbers of all three neutrino flavors -- electron, muon, and tau -- and polarizations -- referred to as ``neutrinos'' (left-handed) and ``antineutrinos'' (right-handed). At low neutrino energies, IMB and Kamiokande-II are only sensitive to the daughter electron antineutrinos and in this case only $1/6$ of the flux in \cref{eq:flux_gamma_nu} is accessible.

We perform two complementary analyses, dubbed ``low energy'' and ``high energy''.
For the former, we investigate if decaying sterile neutrinos can significantly enhance the number of reported  events in the energy interval $E_\nu \in[10,50]$~MeV. We set the time windows to be $\Delta t < 13~(6)$ s for Kamiokande-II (IMB) since these are the time windows inside which SN1987A neutrinos were detected. 
The event rates are given by
\begin{align}
N_\nu^{\text{BSM}}=N_\text{tgt} \int dE_\nu \frac{dN_\nu}{dE_\nu dA}(E_\nu) \,\, \sigma_{\rm IBD}(E_\nu)\,\varepsilon(E_\nu)\,,
\label{eq:event_rates}
\end{align} 
where $N_\text{tgt}$ is the number of hydrogen atoms in the detector -- Kamiokande-II (IMB) contains 2.14 (6.8) kton of water \cite{Fiorillo:2022cdq}, $\sigma_{\rm IBD}(E_\nu)$ is the cross section for inverse beta decay and $\varepsilon(E_\nu)$ is the detection efficiencies~\cite{Fiorillo:2022cdq}. We use \cref{eq:poissonian} and take the number of SN events from our simulation as a proxy for $N_\text{obs}$ to keep the estimates of $N_\text{exp}$ and $N_\text{obs}$ on equal footing. In \cref{eq:poissonian}, we sum the contributions of the two experiments under consideration.
 In this low-energy analysis, we find that the exclusion limits are at most comparable with the cooling limit. 

Regarding the high-energy analysis, restricted to $E_\nu > 70$ MeV, we make use of the fact that no significant excess of neutrino events was reported by either Kamiokande-II \cite{Kamiokande-II:1987idp} or IMB \cite{Bionta:1987qt} in a detailed search for neutrino events inside a two-day interval surrounding the observation of the burst. Atmospheric neutrinos are the main source of background in this analysis. We estimate the number of atmospheric neutrino events inside $\Delta t<1$ day. Further, taking into account that, for $E_\nu \gtrsim$ 70 MeV, the cross sections for both $\nu_e$ and $\bar{\nu}_e$ scattering on oxygen dominate over that for inverse beta decay~\cite{Fiorillo:2022cdq}, we again perform an analysis following \cref{eq:event_rates,eq:poissonian}. 
For $N_\text{obs}$, we use the reported atmospheric neutrino background from IMB (2 events/day), and for Kamiokande-II we rescale the IMB number by the ratio of the fiducial masses of the detectors, a factor of $0.32$. 
The excluded region, combining both IMB and Kamiokande-II data, is depicted in \cref{fig:exclusion} (cyan). We observe that this neutrino limit is stronger than the cooling bound by, roughly, a factor of two. 
Unlike the photon bound (blue shaded), the neutrino limit extends to higher $M_N$ values since daughter neutrinos can come from sterile neutrinos  decaying deep inside the photosphere (still outside of the neutrinosphere). 
The neutrino limit on $d$ is mostly independent from  $M_N$ for $M_N\sim [0.1,100]$~MeV. The limit weakens for lower values of $M_N$, when $\Delta t$ fails to fulfill our selection criteria.
 
The near-future neutrino experiments DUNE \cite{DUNE:2015lol} and Hyper-Kamiokande \cite{Hyper-Kamiokande:2018ofw} are expected to identify several thousands of neutrino events from a future galactic SN. 
For a hypothetical future galactic SN at $D_{\rm SN} =\unit[10]{kpc}$, we estimate the sensitivity of the equivalent of the high-energy analysis for these next-generation experiments. For Hyper-Kamiokande, the strategy is very similar to that of Kamiokande-II; we simply rescale both the new physics and atmospheric background events according to the ratio of fiducial masses. 
In the case of DUNE, for $E_\nu<100$ MeV, we use the antineutrino cross sections on argon presented in \cite{GilBotella:2003sz}. For higher energies (up to 400~MeV), we use  GENIE \cite{Andreopoulos:2009rq} to estimate the relevant cross sections. 
The results are depicted in \cref{fig:exclusion} as dashed yellow lines. 
One is sensitive to $d\gtrsim 10^{-14}$\,$\text{MeV}^{-1}$, roughly an order of magnitude improvement with respect to current limits from SN1987A. Notice that the upturn in the sensitivity occurs at larger masses relative to the constraints from Kamiokande-II and IMB. This is because for relatively smaller values of $d$ relatively larger values of $M_N$ are required to ensure the decay length is shorter than the SN distance.

\begin{figure}[t]
	\centering
	\includegraphics[scale=0.55]{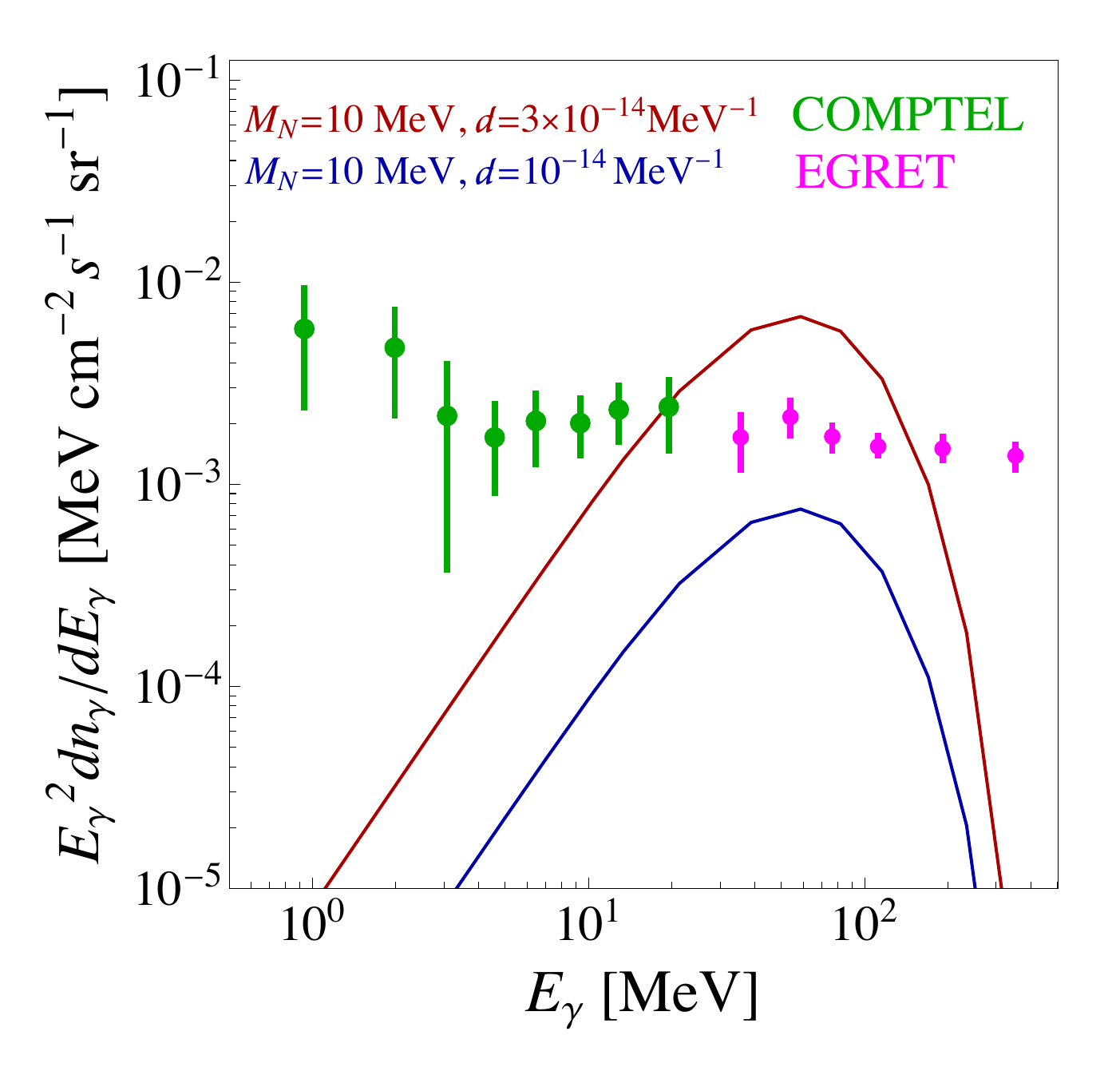} 
	\caption{Diffuse photon flux from sterile neutrino decays compared to the extragalactic photon background measured by COMPTEL \cite{Strong:1994} (green) and EGRET \cite{Hunter:1997} (purple); for more data in this energy regime see \cite{Caputo:2021rux}.}
	\label{fig:diffuse_sepctrum_gamma}
\end{figure}

\textit{\bf{Diffuse $\gamma$-ray and neutrino background ---}}
\label{sec:diffuse}
We also investigate the cumulative effect from all past SN explosions and explore the prospects of detecting the related diffuse $\gamma$-rays and neutrinos from radiative sterile neutrino decays. The number density of sterile neutrinos from all past SN is \cite{Lunardini:2009ya,Caputo:2021rux} 
\begin{align}
\frac{d n_N}{dE}= \frac{c}{4\pi} \int_0^\infty dz (1+z) \, n_\text{cc}'(z) \, \frac{d\mathcal{N}_s}{dE}(E_z)\,,
\label{eq:diffuse}
\end{align} 
where $z$ is the redshift, $n_\text{cc}'$ is the derivative with respect to $z$ of the number of SN per comoving volume, from \cite{Yuksel:2008cu}, and $d\mathcal{N}_s/dE$ is the sterile neutrino spectrum evaluated at $E_z\equiv E(1+z)$. 
To estimate the number density of photons or neutrinos from $N$ decays, we also need the fraction $f_D(z)$ of sterile neutrinos that have decayed by the present time \cite{Caputo:2021rux}. In a nutshell, $(d n_{\gamma,\nu}/dE_{\gamma,\nu})$ is obtained by replacing $d\mathcal{N}_s/dE$ in \cref{eq:diffuse} with $\int_{(1+z)E_{\gamma/\nu}}^\infty dE_z f_D E_z^{-1} (d\mathcal{N}_s/dE)(E_z)$. Diffuse photon spectra for two benchmarks are shown in \cref{fig:diffuse_sepctrum_gamma}. 

We rule out the region of the $d\times M_N$ parameter space where the expected photon energy distribution overshoots the extragalactic differential photon background measured by COMPTEL and EGRET \cite{Caputo:2021rux}, depicted in \cref{fig:diffuse_sepctrum_gamma}. These constraints are shown in \cref{fig:exclusion} (magenta). The shape of the constrained region is different from the constraints associated to a single SN explosion, including the SN1987A constraint, mostly due to effects associated to the cosmological redshift. 

For the diffuse neutrino flux from the radiative decay of sterile neutrinos produced in all past SN, we find it to be unobservable given existing constraints on the model. More details are presented in Appendix \ref{app:diffuse}.

\textit{\bf{Summary and Conclusions ---}}
We considered the neutrino magnetic moment portal between light neutrinos and hypothetical sterile neutrinos with masses $M_N\lesssim 100$ MeV. The new interaction allows for the production of these sterile neutrinos inside SN and mediates their subsequent decays into active neutrinos and photons. If the decay is neither too fast nor too slow, these daughter neutrinos and photons should find their way to the Earth and into our detectors and telescopes. We estimated new constraints  on the magnetic moment portal from the absence of such neutrinos and photons associated to SN1987A and found that these are more powerful probes than the more standard cooling bounds.
We also considered the flux of neutrinos and photons from the radiative decays of sterile neutrinos produced in all previous SN events and showed that constraints from the observed diffuse $\gamma$-ray flux are competitive with those from SN1987A for $M_N \lesssim 1$ MeV. With an eye towards the future, we estimated the sensitivity of next SN explosion, given the capabilities of current and next-generation neutrino detectors and $\gamma$-ray telescopes.




\textbf{Acknowledgements.}
We would like to thank Thomas Janka for providing the data from Garching core-collapse SN simulations in machine-readable form. This work was supported in part by the US Department of Energy (DOE) grant \#de-sc0010143 and in part by the NSF grant PHY-1630782.
Fermilab is managed by the Fermi Research Alliance, LLC (FRA), acting under Contract No.\ DE-AC02-07CH11359. Y.-Y. L is supported by the NSF of China through Grant No. 12047502.


\appendix
\section{Sterile Neutrino Spectrum}
\label{app:integrate}
Given the differential number of sterile neutrinos $\mathcal{N}_s$ 
\begin{align}
\frac{1}{4\pi r^2}\frac{\partial^2}{\partial r\partial t}\left(\frac{d\mathcal{N}_s}{dE_N}\right)= \sigma n_e \frac{d n_\nu}{dE}\,,
\label{eq:simplified}
\end{align}
the energy distribution of the sterile neutrinos is
\begin{align}
\frac{d\mathcal{N}_s}{dE_N}&=\int_0^R  4\pi R'^2 dR' \int_0^t dt'  \,\, \frac{(1+\alpha(t',R'))^{(1+\alpha(t',R'))}}{\Gamma(1+\alpha(t',R'))\, \bar{E}(t',R')} 
\nonumber \\ & \left[\frac{E}{\bar{E}(t',R')}\right]^{\alpha(t',R')}\,\text{Exp}\big[-(1+\alpha(t',R')) \, \frac{E}{\bar{E}(t',R')}\big]\nonumber \\ & n_\nu(t',R')\,\, \sigma(d,M_N,T(t',R'),E) \, n_e(t',R')\,.
\label{eq:sterile_spectrum}
\end{align}
The expression on the right-hand side is to be summed over all flavor species of antineutrinos and neutrinos.  Here, $\alpha$, $\bar{E}$, and the temperature $T$ characterize the energy distribution of the active neutrinos as a function of radius and time. In particular, $\alpha$ \cite{Keil:2002in} parameterizes the deviation of the SN neutrino energy distribution from a Maxwell-Boltzmann one; all $\alpha$-dependent terms in \cref{eq:sterile_spectrum} are building blocks of $f(E)$, see \cite{Brdar:2018zds}. All parameters are evaluated at ``snapshots'' in time and space (provided by SN simulation data) between which we interpolate. The radial integral runs from the center of the star (close to $R'=0$) to its outer layers; in our calculations we stop at $40$~km since the production of  sterile neutrinos is concentrated inside the neutrinosphere ($\sim 30$~km) due to the larger number densities of neutrinos and electrons. The time integration encompasses the neutronization, accretion, and cooling phases of the SN explosion and, for the simulation at hand, data are available until $8.85$~s after core bounce. 

\begin{figure}[t]
	\centering
	\includegraphics[scale=0.56]{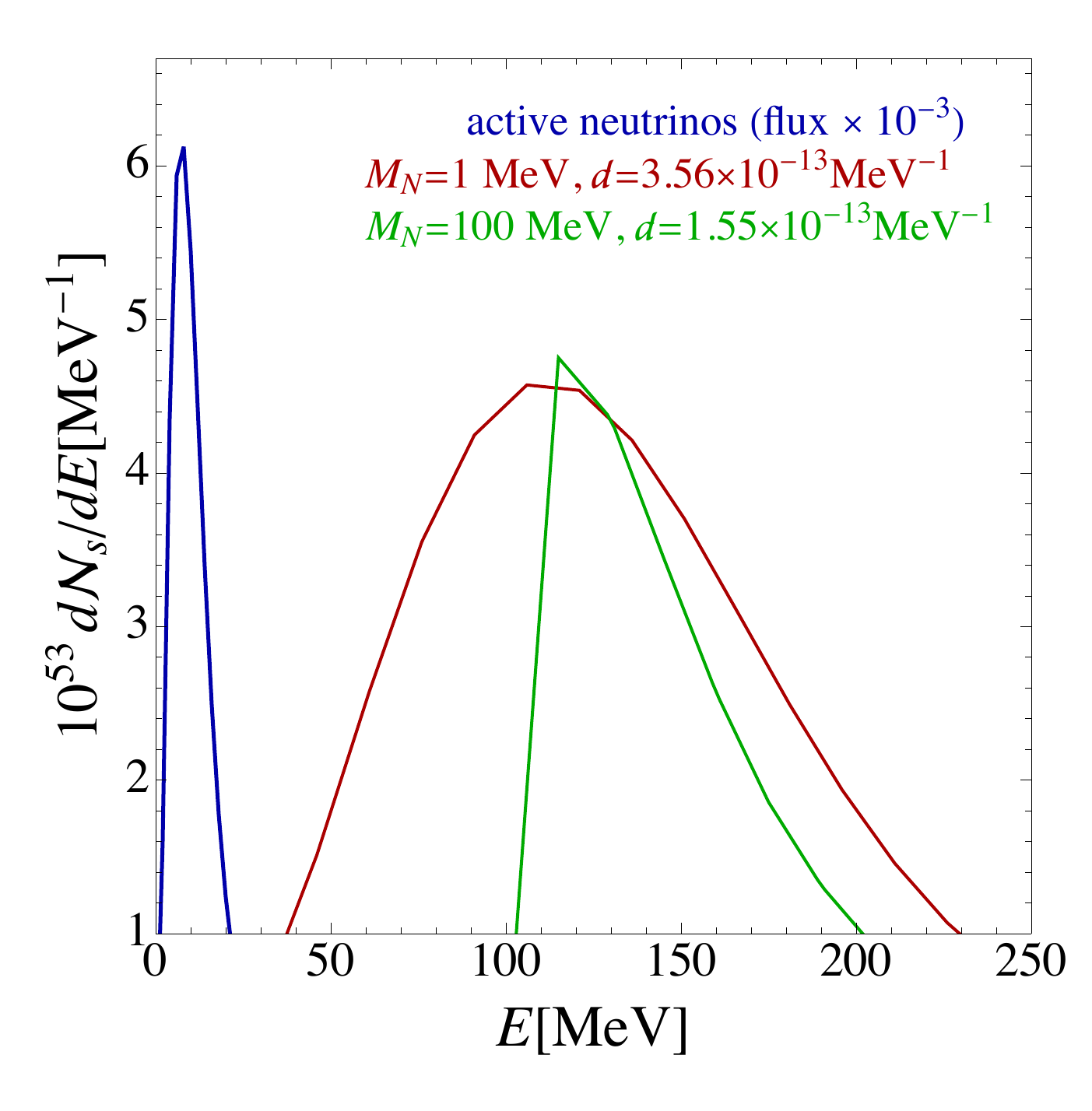} 
	\caption{Thermal active neutrino spectrum (blue) relative to that of  sterile neutrinos (red, green) produced through the magnetic moment portal. See text for details.}
	\label{fig:fluxes}
\end{figure}

We used data associated to the simulation of a $8.8 M_\odot$ progenitor star performed by the Garching group \cite{Huedepohl2010} and do not explore potential uncertainties associated to SN modeling \cite{Bollig:2017lki, Vartanyan:2018iah}. We expect these not to impact our results in a meaningful way. 
\cref{fig:fluxes} depicts sterile neutrino spectra calculated using \cref{eq:sterile_spectrum} for two benchmark points with $M_N=1$ MeV (red) and $M_N=100$ MeV (green) and comparable values of $d$. For comparison, the figure also contains the active neutrino spectrum (summed over all flavors, in blue). The typical sterile neutrino energies (as well as the energies of its decay products) are roughly an order of magnitude larger than the energies of the thermal active neutrinos.

\section{Diffuse Neutrino Background from Sterile Neutrino Decays}
\label{app:diffuse}
\begin{figure}[t]
	\centering
	\includegraphics[scale=0.6]{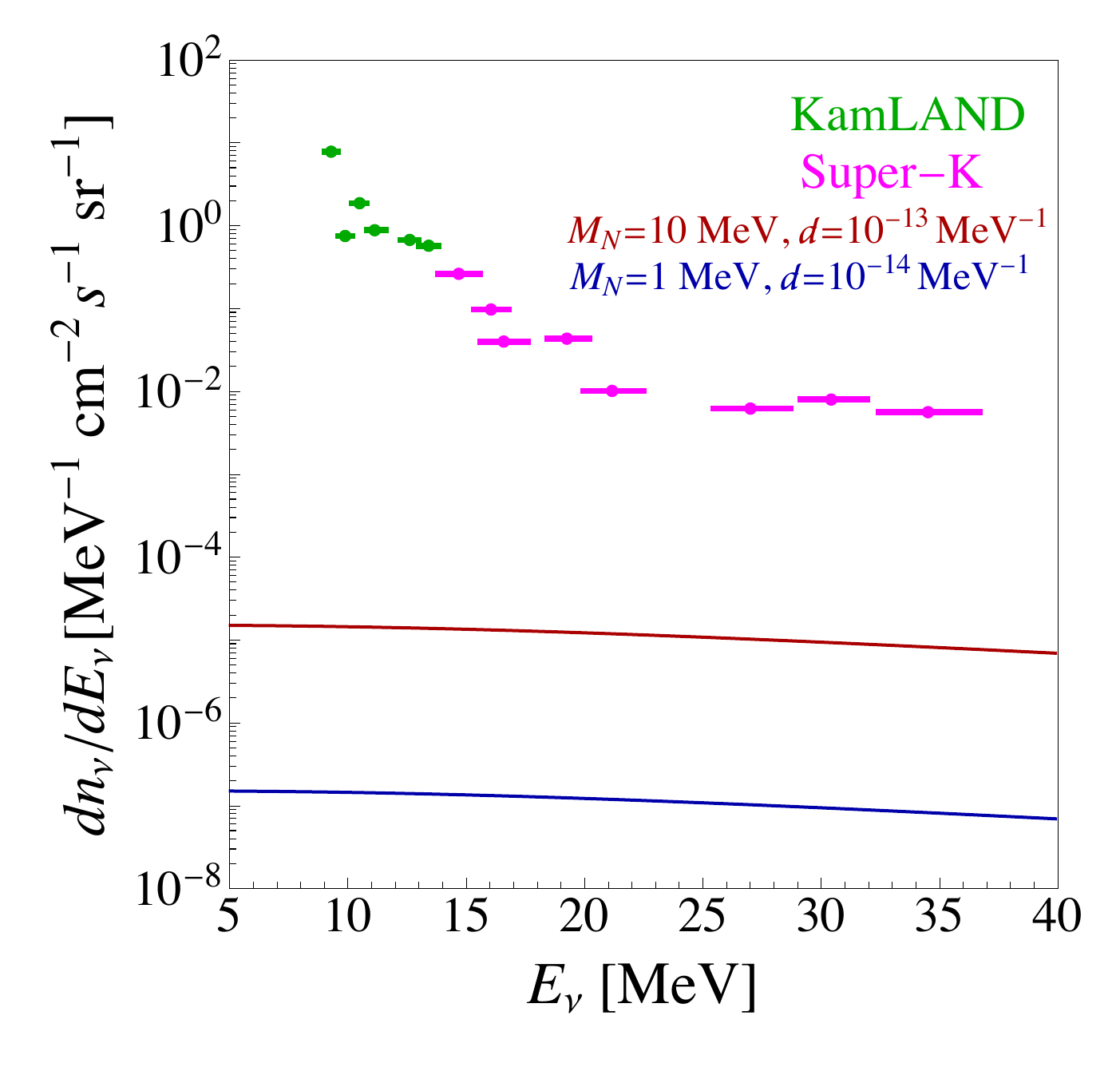} 
	\caption{Diffuse neutrino flux from sterile neutrino decays compared to the flux sensitivity of KamLAND and Super-Kamiokande.}
	\label{fig:diffuse_spectrum_nu}
\end{figure}

For the detection of active neutrinos from the decay of sterile neutrinos hypothetically produced inside all past SN explosions, we concentrate on the detection of electron antineutrinos via inverse beta decay in water Cherenkov detectors which is the most sensitive channel. We consider results from \cite{Super-Kamiokande:2021jaq} and utilize their 90\% CL upper limit on the diffuse $\bar{\nu}_e$ flux (see Fig.~25 in \cite{Super-Kamiokande:2021jaq}, where previous results from Super-Kamiokande \cite{PhysRevD.85.052007} and KamLAND \cite{KamLAND:2021gvi} are also summarized). In \cref{fig:diffuse_spectrum_nu}, we compare the fluxes expected from two benchmark points with the existing upper limits. As can be inferred from the figure, for values of $d\simeq 10^{-13}$\, $\text{MeV}^{-1}$, the BSM-induced neutrino diffuse flux is significantly smaller than the existing upper limits. These values of $d$ roughly correspond to the neutrino limits from the SN1987A energy loss argument so current bounds from the diffuse neutrino background are not competitive. For $E_\nu\gtrsim 40$ MeV, there are no upper limits on the diffuse flux in the literature; for these and higher energies, the atmospheric neutrino background is relatively large.

\bibliography{refs}

\end{document}